# Observing atom bunching by the Fourier slice theorem

A. Blumkin,[1] S. Rinott,[1] R. Schley,[1] A. Berkovitz,[1] I. Shammass,[2] and J. Steinhauer[1]

[1]*Department of Physics, Technion—Israel Institute of Technology, Technion City, Haifa 32000, Israel*

[2]*Department of Condensed Matter Physics, Weizmann Institute of Science, Rehovot 76100, Israel*

By a novel reciprocal space analysis of the measurement, we report a calibrated *in situ* observation of the bunching effect in a 3D ultracold gas. The calibrated measurement with no free parameters confirms the role of the exchange symmetry and the Hanbury Brown-Twiss effect in the bunching. Also, the enhanced fluctuations of the bunching effect give a quantitative measure of the increased isothermal compressibility. We use 2D images to probe the 3D gas, using the same principle by which computerized tomography reconstructs a 3D image of a body. The powerful reciprocal space technique presented is applicable to systems with one, two, or three dimensions.

In a gas of non-interacting bosons or fermions, the fluctuations and correlations are increased or decreased relative to the case of randomly located particles [1]. These bunching and antibunching effects are due to the exchange symmetry in the many-body wavefunction, and are the spatial versions of the Hanbury Brown-Twiss (HBT) effect [2]. However, the fluctuations can also be affected by interactions [3-6], light-assisted collisions [7,8], or losses [9,10], with no direct relation to the HBT effect. The role of the HBT effect could potentially be verified by the anisotropy of the correlation function during ballistic expansion [11]. Another method of demonstrating that the bunching is truly due to the HBT effect is by a calibrated measurement of the magnitude of the effect [12], since the exchange symmetry gives a correlation function of two or zero for bosons or fermions, respectively.

The bunching or antibunching effects for atoms have been observed in a thermal beam [13], a pseudothermal beam [14], a Mott insulator [15,16], a 1D Bose gas [17,18], a 3D Fermi gas [19-22], and a 3D Bose gas [19,23-26]. In the 3D Bose case, a gas of metastable helium atoms was released from its trap and allowed to fall onto a microchannel plate detector [19,24,25]. The arrival of the atoms at the detector showed the bunching effect. For a 3D Bose gas of $^{87}$Rb atoms, the three-body recombination rate was used to study third-order correlations [23], or the gas was probed temporally by an electron beam [26]. Here we study a 3D Bose gas of $^{87}$Rb atoms by a very different technique. We image the atoms *in situ* and observe the bunching spatially in the 2D image. This requires overcoming two major technical problems. Firstly, the thickness of the sample perpendicular to the imaging direction can wash out the apparent fluctuations. This problem is unique to the case of a 3D gas. Secondly, the limited resolution of the imaging system presents a challenge, regardless of the dimensionality. We avoid these difficulties through a reciprocal space analysis of the *in situ* images. This allows us to make a fully-calibrated observation of the bunching effect in an *in situ* 3D Bose gas, with no free parameters [12]. This calibrated measurement confirms the role of the exchange symmetry of the many-body wavefunction.

The spatial version of the HBT effect is typically observed by measuring the two-body correlation function $g^{(2)}(r)$, where $r$ is the distance between any two points in the homogeneous system [19,24]. For an ideal homogeneous Bose gas, the correlation function is given by [1]

$$g^{(2)}(r) = 1 + \frac{1}{(2\pi)^6 n^2}\left|\int d\mathbf{k}\, n_k e^{i\mathbf{k}\cdot\mathbf{r}}\right|^2 \qquad (1)$$

where $n_k = \left[e^{(\hbar^2 k^2/2m - \mu)/k_B T} - 1\right]^{-1}$ is the Bose distribution, $m$ is the atomic mass, $\mu$ is the chemical potential, $T$ is the temperature, and $n$ is the density. For $r$ much greater than the thermal wavelength $\lambda = \sqrt{2\pi\hbar^2/mk_B T}$, the contributions from the various $n_k$ average to zero, and $g^{(2)}(r)$ approaches unity, as for a random distribution of particles. For $r \ll \lambda$ however, $g^{(2)}(r)$ approaches two. Fig. 1(b) shows (1) for the temperature range studied in this experiment. In order to observe the HBT effect, one should observe that $g^{(2)}(r)$ exceeds unity. As seen in the figure, a spatial resolution of less than 0.2 μm is thus required. However, the shaded region of the figure shows the regime corresponding to the resolution of our imaging system. Thus, the HBT effect is far from resolvable.

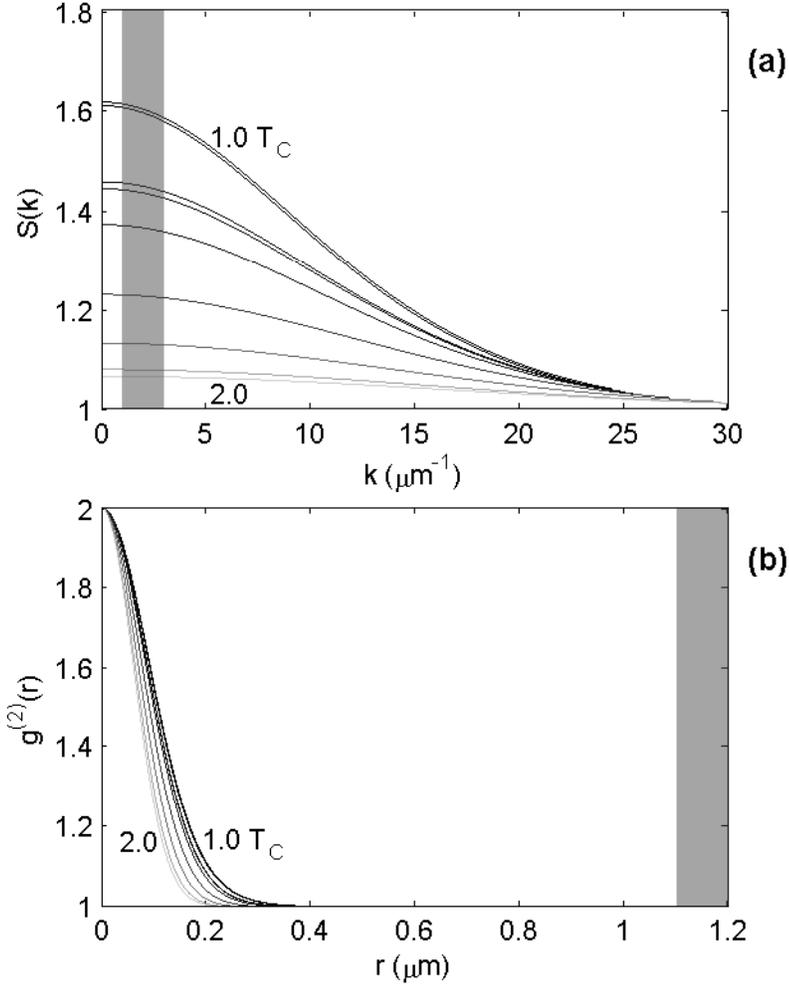

FIG. 1. The spatial Hanbury Brown-Twiss effect as studied in $k$-space versus $r$-space. The temperatures and densities shown correspond to the measurements of Fig. 3. (a) The static structure factor at various temperatures. The deviation from unity corresponds to the HBT effect. The shaded region indicates the $k$-window used in the present experiment. (b) The two-body correlation function. The deviation from unity corresponds to the HBT effect. The shaded region indicates the regime within the resolution of the imaging system.

These measurement limitations can be partially circumvented by studying the density fluctuations in small but spatially resolved sub-volumes of the gas, as performed for a 1D

Bose gas [17,18] and a Fermi gas [21,22]. These fluctuations are given by $\langle\delta N_s^2\rangle/N_s = 1 + n \int d\mathbf{r}[g^{(2)}(r) - 1]$, where $N_s$ is the number of atoms in the sub-volume.

The short length scale of the correlations in position space becomes an advantage in reciprocal space ($k$-space). The fact that $g^{(2)}(r) - 1$ in Fig. 1(b) is such a narrow function of $r$ implies that its Fourier transform is a broad, almost constant function of the wavenumber $k$, similar to white noise, as seen in Fig. 1(a). The Fourier transform is expressed by the static structure factor

$$S(k) = 1 + n \int d\mathbf{r}[g^{(2)}(r) - 1]e^{i\mathbf{k}\cdot\mathbf{r}}. \tag{2}$$

Thus, $S(k)$ differing from unity is equivalent to the HBT effect, in which $g^{(2)}(r)$ differs from unity. $S(k)$ gives the spectrum of the density fluctuations in $k$-space [27], which can be directly measured from the images by the relation [28]

$$S(k) = \frac{1}{N}[\langle|\rho_\mathbf{k}|^2\rangle - |\langle\rho_\mathbf{k}\rangle|^2], \tag{3}$$

where the averages are taken over the ensemble of images, $\rho_\mathbf{k}$ is the Fourier transform of the density, and $N$ is the number of atoms. The static structure factor (3) is not the same as the zero-temperature static structure factor, which is measured by observing the response of the gas to a Bragg pulse [29]. The zero-temperature static structure factor is insensitive to the density fluctuations [29], and is therefore not relevant for this work. By (2), the fluctuations $\langle\delta N_s^2\rangle/N_s$ are given by $S(k = 0)$. Since $S(k)$ is so broad, we do not need to measure at precisely $k = 0$. We can make the observation at any convenient spatial frequency, thus bypassing the resolution limitation, and avoiding various sources of noise. As seen in Fig. 1(a), the spectrum is much broader than the $k$-window accessible by our imaging system, which is indicated by the shaded region. By measuring $S(k)$ in the $k$-window, Fig. 1(a) shows that we are essentially measuring $S(k \approx 0)$. The HBT effect ($S(k)$ exceeding unity) is seen to be strongest at these long wavelengths, so the limited resolution of the measurement no longer presents a problem. Note that it is impossible to measure for $k$ precisely zero in our closed system, since $S(k = 0)$ gives the fluctuations in the total number of particles in the gas.

The second problem overcome by our $k$-space measurement is the integration of the density in the imaging direction, since the integration in position space gives a slice in $k$-space. Fig. 2(b) shows the density $\rho(\mathbf{r})$ integrated in the $z$-direction, perpendicular to the image. By the Fourier slice theorem, the 2D Fourier transform gives a slice in $k$-space, $\rho(k_x, k_y, k_z = 0)$. We thus obtain the slice of the static structure factor $S(k_x, k_y, k_z = 0)$, by (3). Thus, the integration does not wash out the effect. By merely taking the Fourier transform of each image, we therefore overcome both of the imaging difficulties, and make a calibrated observation of the bunching effect.

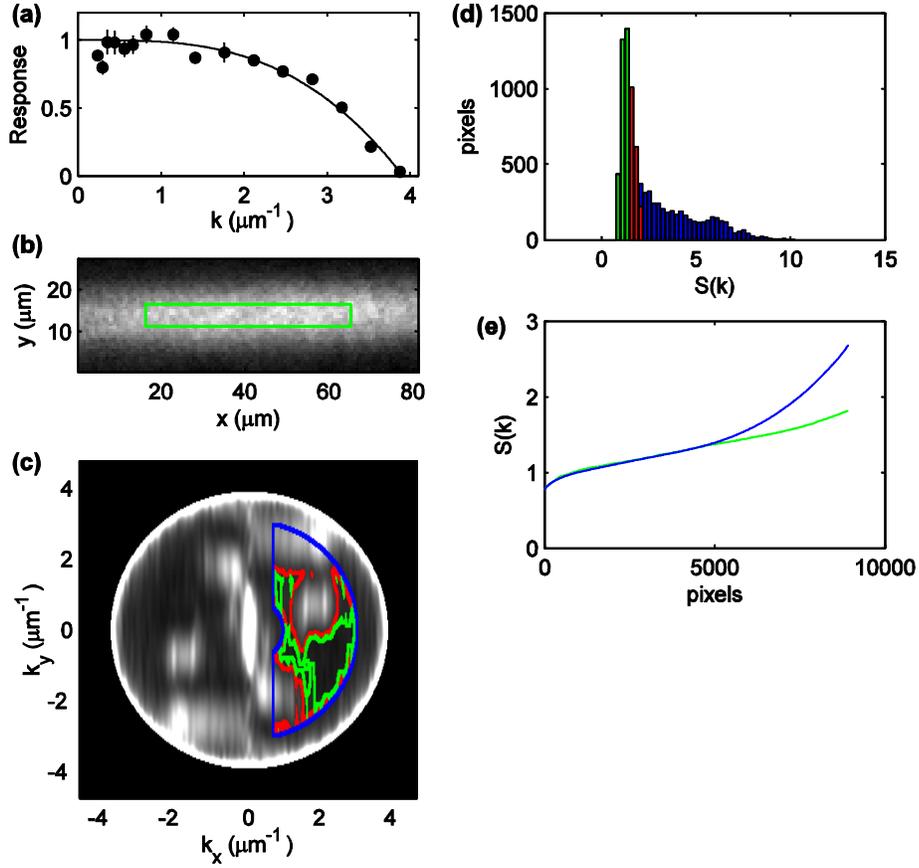

FIG. 2. Measuring the static structure factor. (a) The response of the imaging system. The measured points are the ratio between the measured and expected response of a condensate to short Bragg pulses. The solid curve is a polynomial fit, which is used as the calibration. (b) *In situ* image of the Bose gas at $T = 1.5\, T_c$. The Fourier transform is computed within the green rectangle. (c) The static structure factor in the $k_z = 0$ plane. The average over all temperatures is shown. The area within the green (red) curve is the minimal (maximal) "clean window" used to compute $S(k)$. (d) Histogram of $S(k)$ for the pixels within the blue curve of (c). The green bars correspond to the area outlined in green in (c). The red+green bars correspond to the larger area outlined in red in (c). (e) The mean (blue) and the median (green) of the left section of the histogram in (d), as a function of the number of pixels included in the section. 5000 pixels corresponds to the red+green bars in (d).

Plugging (1) into (2), $S(k)$ can be written as [1]

$$S(k) = 1 + \frac{1}{(2\pi)^3 n} \int n_{\mathbf{k'}} n_{\mathbf{k'+k}} d\mathbf{k'}. \qquad (4)$$

This expression gives us a different perspective on the HBT effect ($S(k)$ exceeding unity). It results from matter wave interference between pairs of populations $n_{\mathbf{k'}}$ and $n_{\mathbf{k'+k}}$. The resulting interference fringes increase the fluctuation spectrum $S(k)$ above unity. As $T$ decreases toward the critical temperature $T_c$ for Bose-Einstein condensation, the small-$k$ populations increase, giving an increase in the HBT effect. Indeed, $S(k \approx 0)$, the quantity probed in this experiment, is a function of the phase-space density $n\lambda^3$ only.

We performed a preliminary experiment to determine the response of the imaging system as a function of $k$ [30]. We created counterpropagating phonons in a Bose-Einstein condensate via short Bragg pulses. We compared the apparent $\rho_\mathbf{k}$ in the image with the simulated $\rho_\mathbf{k}$. The ratio between these values gives the response shown in Fig. 2(a), where the long-wavelength response has been brought to unity by an overall factor. The solid curve is a polynomial fit, which is used as the response function which calibrates the current experiment.

In this experiment, the atomic cloud consists of approximately $7 \times 10^5$ $^{87}$Rb atoms in the $F = 2$, $m_F = 2$ state, confined in a harmonic magnetic potential with radial and axial frequencies of 224 Hz and 26 Hz, respectively. This Bose gas is cooled by RF evaporation to a temperature greater than the critical temperature at which a condensate appears in the center of the trap, given by $T_c = 390$ nK. The cloud is then imaged by phase contrast imaging, as shown in Fig. 2(b). We employ a short 2 μs imaging pulse and a relatively small detuning of 210 MHz. The latter enhances the signal for the low optical density cloud. An ensemble of usually 20 images is collected at a given temperature. To find $S(k_x, k_y, k_z = 0)$ by (3), we compute the 2D Fourier transform of each image, within the green rectangle of Fig. 2(b). We then make two corrections. Firstly we subtract off the shot noise due to the camera and imaging laser. The shot noise is determined from the Fourier transform for $k > 4$ μm$^{-1}$, since there is no atomic signal for these values of $k$, as seen in Fig. 2(a). Secondly, we divide by the square of the response function of the imaging system shown in Fig. 2(a). Fig. 2(c) shows the resulting $S(k_x, k_y, k_z = 0)$. The region within the circle is resolved by the imaging system. In order to clearly find the imaging artifacts, Fig. 2(c) has been averaged over all of the temperatures studied. The central white spot corresponds to the overall shape of the cloud. The second term in (3) removes most of this artifact, but the 1% which remains is the strongest feature in Fig. 2(c). In order to avoid this artifact and yet remain well within the resolution of the imaging system, we confine our study to the region outlined in blue. This region contains $k$-values ranging from 1 μm$^{-1}$ to 3 μm$^{-1}$. This region contains flat gray areas (the desired atomic fluctuations), as well as white peaks corresponding to sinusoidal fringes in the image. The latter result from fluctuations in the coherent light of the imaging laser. These fringes are too weak to be detected by eye in the image of Fig. 2(b). While one can clearly differentiate the flat gray signal from the white imaging peaks by inspecting Fig. 2(c), we distinguish them by studying the histogram of the pixels within the blue curve, as shown in Fig. 2(d). From this histogram we estimate the maximal and minimal useful flat gray areas. The minimal area is indicated by the green regions of Figs. 2(c) and 2(d), and the maximal area is indicated by the combined red and green regions. The maximal area is 1.6 times larger than the minimal area. In general, the flat gray area corresponds to the left region of the histogram. The flatness of this area is clear from the very steep left edge of the histogram. The minimal green area is just wide enough to contain the left peak of the histogram. In order to find the maximal area, we assume that the noise in the flat gray area has a symmetric distribution, implying that the mean is approximately equal to the median. This is clearly not the case for the entire histogram, with its "tail" extending to the right, corresponding to the imaging peaks. Fig. 2(e) shows the mean and the median of the distribution, as a function of the number of pixels included in the maximal area. The mean and median are approximately equal, as long as the area is 5000 pixels or less. We thus take 5000 pixels as the maximal area.

Averaging $S(k)$ over the pixels in the minimal or maximal areas, $S(k \approx 0)$ is obtained. This is shown in Fig. 3(a) as a function of $T/T_c$. As the temperature decreases toward $T_c$, bunching is clearly seen, in that $S(k \approx 0)$ increases above unity. As discussed above, the plotted quantity is approximately equal to $\langle \delta N_s^2 \rangle / N_s$ and $1 + n \int [g^{(2)}(r) - 1] d\mathbf{r}$, and is a measure of $n\lambda^3$.

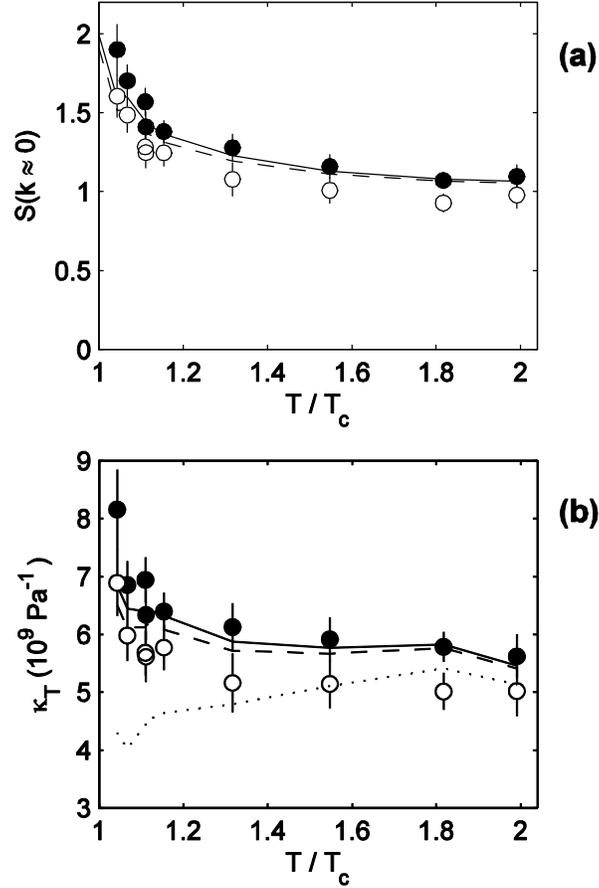

Fig. 3. Calibrated observation of the bunching effect, with no free parameters. The filled (open) circles correspond to the maximal (minimal) observation area. The solid curve is the ideal Bose gas model. The dashed curve includes interactions. The error bars indicate the standard error of the mean. (a) The static structure factor for long wavelengths. Values above unity correspond to the bunching effect. (b) The isothermal compressibility. The dotted curve indicates $\kappa_T$ of a classical ideal gas, for which $S(k \approx 0)$ is unity.

The temperature $T$ is determined from Fig. 2(b) by a fit of a semiclassical density profile obtained within the Hartree-Fock approximation [28,31]. This fit also yields the average density within the green rectangle of Fig. 2(b). This temperature and density are used to compute the theoretical $S(k \approx 0)$ for an ideal Bose gas, by averaging (4) within the $k$-window shown in Fig. 1(a). The result is indicated by the solid curve of Fig. 3(a). The small kinks in the curve reflect the variations in the experimental density. The quantitative agreement between the experiment and the ideal Bose gas model is seen to be very good, with no free parameters. The RMS deviation of the experiment from the model is 10% for the minimal area, and 8% for the maximal area.

We also show the small predicted reduction in the bunching due to the repulsive interactions, indicated by the dashed curve of Fig. 3(a). $g^{(2)}(r)$ of Fig. 1(b) has length scale $\lambda$, but the interactions suppress the correlations for distances on the order of the $s$-wave scattering length $a$. Thus, the importance of the interactions is quantified by the ratio $a/\lambda$ [32], which is 0.02 for our experiment. The interacting curve is calculated by inserting $g^{(2)}(r) = 1 + 2a^2/r^2 + \left[g^{(2)}(r)_{\text{ideal}} - 1\right](1 - 4a/r)$ in (2) [33].

The enhanced long-wavelength fluctuations of the bunching effect, quantified by $S(k \approx 0)$, imply that the gas is readily compressible. This is expressed by the fluctuation-dissipation theorem, which gives the relation $\kappa_T = S(k = 0)/nk_BT$, where the isothermal compressibility is defined by $\kappa_T \equiv n^{-1}\partial n/\partial P$, and $P$ is the pressure [28,34]. Fig. 3(b) shows the resulting values of $\kappa_T$. The exceedingly large values shown are similar to the compressibility of a Bose-Einstein condensate in the same confining potential. Fluctuations were also used to measure $\kappa_T$ in a strongly interacting Fermi gas [35]. $\kappa_T$ can also be measured by observing the overall density profile in a known confining potential [16], as applied to a Mott insulator [16], a Fermi gas [21,36], and a 2D Bose gas [37].

In conclusion, we have made a fully calibrated *in situ* measurement of the bunching effect in a 3D Bose gas. The result is in very good agreement with the model of a homogeneous, ideal Bose gas. This confirms the role of the exchange symmetry in the effect, which gives the factor of two in the correlation function. The result also suggests that interactions, light-assisted collisions, and losses play a negligible role. Furthermore, the bunching effect gives a measurement of the isothermal compressibility, which is found to be 15 orders of magnitude larger than the compressibility of air. By measuring closer to the phase transition, the critical exponent for the isothermal compressibility could be extracted. The powerful $k$-space technique presented could be used to study the correlations in a variety of systems, including optical lattices.

We thank N. Pavloff, L. I. Glazman, G. V. Shlyapnikov, and I. Zapata for helpful conversations. This work was supported by the Israel Science Foundation.